\documentclass[twoside,11pt]{article}

\usepackage{jmlr2e}

\firstpageno{1}

\usepackage[utf8]{inputenc} 
\usepackage[T1]{fontenc}    
\usepackage{hyperref}       
\usepackage{url}            
\usepackage{amsfonts}       

\usepackage{amsmath}
\usepackage{mathtools}
\usepackage{mathrsfs}
\usepackage{graphicx}
\usepackage{natbib}
\usepackage{enumitem}
\usepackage{caption}
\usepackage{subcaption}
\usepackage{cleveref}

\usepackage{subcaption}

\DeclareCaptionLabelFormat{opening}{(#2)}

\begin{document}

\title{Recurrent neural network models for working memory of continuous variables: activity manifolds, connectivity patterns, and dynamic codes}

\author{\name Christopher J. Cueva$^*$ \email ccueva@gmail.com \\
    \addr Department of Brain and Cognitive Sciences \\
       MIT, Cambridge, MA, USA
      \AND
      \name Adel Ardalan\thanks{Equal contribution.} \email adel.ardalan@princeton.edu \\
      \addr Princeton Neuroscience Institute \\
      Princeton University, Princeton, NJ, USA
      \AND
      \name Misha Tsodyks \email mtsodyks@gmail.com \\
      \addr Simons Center for Systems Biology \\
      School of Natural Sciences \\
      Institute for Advanced Study \\
      Princeton, NJ, USA
      \AND
      \name Ning Qian \email nq6@columbia.edu \\
      \addr Department of Neuroscience, Zuckerman Institute \\
      Department of Physiology \& Cellular Biophysics\\
      Columbia University, New York, NY, USA
      }

\maketitle

\begin{abstract}
Many daily activities and psychophysical experiments involve keeping multiple items in working memory. When the items take continuous values (e.g., orientation, direction, contrast, length, weight, loudness) they must be stored in a continuous structure of appropriate dimensions. We investigate how such a structure might be represented in neural circuits by training recurrent networks to report two previously flashed stimulus orientations. We find that the activity manifold for the two orientations resembles a Clifford torus. Although a Clifford torus and a standard torus (the surface of a donut) are topologically equivalent, they have important functional differences. A Clifford torus treats the two orientations equally and keeps them in orthogonal subspaces, as demanded by the task, whereas a standard torus does not. We further find that the Clifford-torus-like manifold is realized by two different sets of locally-excitatory/globally-inhibitory connectivity patterns. Moreover, in addition to attractors that store information via persistent activity, our networks also use a dynamic coding scheme such that many units change their tuning to prevent the new sensory input from overwriting the previously stored one. We argue that such dynamic codes are generally required whenever multiple inputs enter a memory system via shared connections. Finally, we apply our framework to a human psychophysics experiment in which subjects reported two remembered orientations. We demonstrate that not all RNNs reproduce human behavior. By varying the training conditions of the RNNs, we test and support the hypothesis that human behavior is a product of both neural noise and reliance on the more stable and behaviorally relevant memory of the ordinal relationship between the two orientations. This suggests that suitable inductive biases in RNNs are important for uncovering how the human brain implements working memory. Together, these results offer an understanding of the neural computations underlying a class of visual decoding tasks, bridging the scales from human behavior to synaptic connectivity.
\end{abstract}

\section{Introduction}

Humans can keep a few items of interest in their working memory and recall them a little later \citep{Ma2014}. A typical example would be to read off a few measurements from a ruler and then jot them down. Many psychophysical studies are similarly designed to test working memory \citep{Miller1956,Cowan2001}. When the items are discrete in nature (e.g., words), they could be stored in point attractors of neural circuits \citep{Hopfield1984}. In contrast, if the items take continuous values (e.g., orientations) \citep{Bae2015,Bae2017,Ding2017,Luu2020}, then they would have to be stored in a continuous structure that could represent any of the possible values. Indeed, a line or ring attractor has been proposed to store an item whose value varies along one dimension \citep{Zhang1996,Compte2000,Machens2005,Itskov2011}. Conceptually, this approach could be readily extended to storing multiple items of continuous values. 

There are, however, a few key questions regarding how a neural circuit of working memory may store multiple items of continuous values. Consider the typical psychophysical task of remembering two successively presented stimulus orientations \citep{Bae2017,Ding2017}. Since a 1D structure of neural activities (such as a ring attractor) is needed to represent any value of each orientation, a manifold of (at least) 2 dimensions is required to store any combination of the two orientations. It is, however, unclear which specific 2D manifold should be used and how the choice of such a manifold depends on the task.  Another question is that given an appropriate activity structure for storing multiple continuous values, what connectivity patterns among the memory units could generate the desired activity structure. Additionally, biologically relevant variables are often encoded by broadly tuned cells. For example, orientation tuning curves of V1 neurons have a full width at half height of about 40 degrees \citep{Schiller1976}. If two stimulus orientations differing by a few degrees are presented successively at the same location \citep{Ding2017}, then they provide very similar inputs, via the same set of connections, to the same set of the memory units. How, then, does the system prevent the memory of the first orientation from being overwritten by the arrival of the second input? 

We investigated these questions by training recurrent neural networks (RNNs) to store two successively flashed orientations, and by analyzing the activity and connectivity patterns of the trained networks. RNNs are a natural choice since the recurrent connections are needed to keep the stimuli in memory after their disappearance. Given the periodic nature of each orientation, one might expect that the activity manifold for storing any two orientations would resemble a standard torus (the surface of a donut), a curved 2D surface embedded in 3D. Instead, we found that the activity manifold is more like a Clifford torus, a flat 2D surface embedded in 4D. Although there are smooth, one-to-one mappings between them, the two types of tori differ in their metric properties. In particular, a Clifford torus treats the two orientations equally and keeps them in orthogonal subspaces, which prevents interference as demanded by the task, whereas a standard torus does not. We further found that the Clifford-torus-like manifold of the memory activities is realized by  two different sets of connections that exhibit locally excitatory and globally inhibitory connectivity patterns in the orientation domain. Moreover, the units storing the memory of the first orientation change their tuning over time \citep{Masse2020,Panichello2021,Wan2021}, transitioning from one attractor-like structure to another, to prevent the memory from being overwritten by the second orientation. Dynamic codes like this could be a general solution to the overwriting problem whenever there is significant overlap among multiple inputs to a memory system \citep{Rademaker2019}.

Finally, we considered the specific study of \cite{Ding2017} in which subjects reported two remembered line orientations and (implicitly) their ordinal relationship (whether the second orientation is clockwise or counterclockwise from the first). 
In this paradigm, absolute orientations of individual lines and their ordinal relationship are the lower- and higher-level features, respectively. 
Ding et al.\ showed that their data contradict the widely assumed low-to-high-level decoding hierarchy which decodes the absolute orientations first, and then compares them to determine the ordinal relationship. 
Instead, they explained their data with a high-to-low-level decoding hierarchy that uses the ordinal relationship to constrain the decoding of the absolute orientations in working memory. Ding et al.\ argued that the high-to-low-level decoding is advantageous because working memory is noisy and so the brain should prioritize the more stable discrete/ordinal memory to constrain the less stable continuous memories of orientations. In addition, higher-level more categorical memories should be prioritized because they are often more behaviorally relevant \citep{Peelen2014,Ekman2003}.
We reasoned that if this is true then not all RNNs trained to implement the psychophysics task of Ding et al., by reporting two absolute orientations and their ordinal relationship, should display the behavioral biases of humans. We hypothesized that injecting noise into RNNs during memory periods helps the networks to learn the high-to-low-level decoding scheme, and our simulations confirmed this hypothesis. 
We also found that the activity manifolds and connectivity patterns of the trained networks are similar to those discussed above. But because the outputs now include the ordinal relationship between the two absolute orientations, the subspaces for the two orientations deviate somewhat from orthogonality, producing a slightly deformed Clifford torus as the activity structure of the memory units. Overall, our results help understand how recurrent neural networks store multiple continuous variables, and shed light on possible neural mechanisms underlying many psychophysical and daily tasks that require working memory.

\begin{figure}[h]
\vspace{-.05in}
\begin{subfigure}{0em} \phantomsubcaption{} \label{fig:taskandarchitectureA} \end{subfigure}
\begin{subfigure}{0em} \phantomsubcaption{} \label{fig:taskandarchitectureB} \end{subfigure}  
\centering
\includegraphics[trim={0cm 12.3cm 1.3cm 0cm},clip,keepaspectratio,width=1\linewidth]{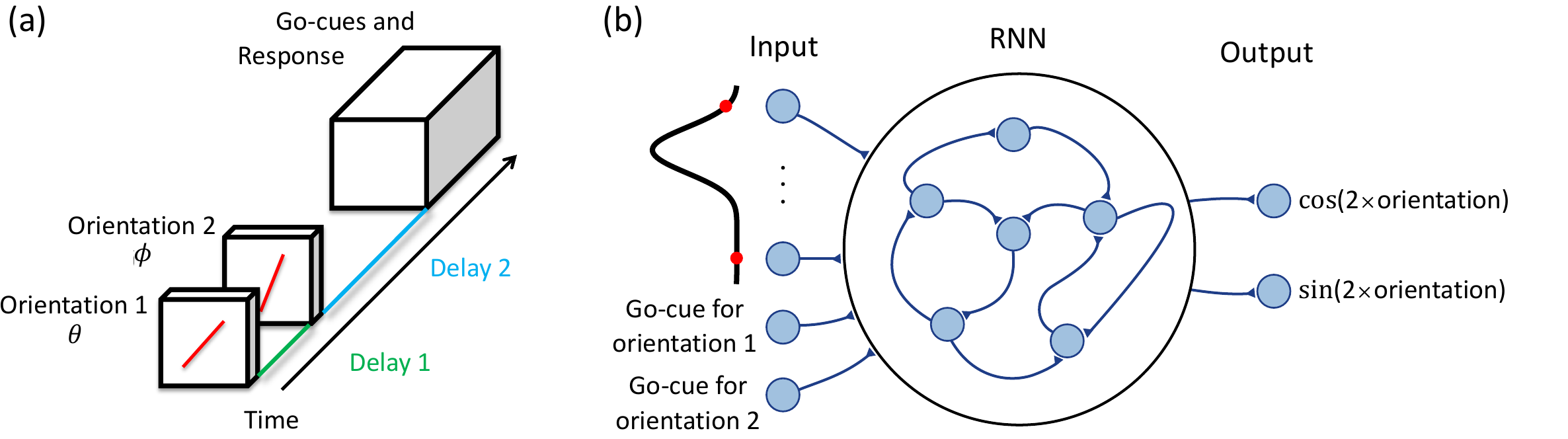}
 \caption[RNN inputs and outputs]{RNN task and architecture. \textbf{(a)} The RNN was trained to remember the orientations of two lines and then report these orientations sequentially when prompted with go-cues. Each input line was followed by a delay. \textbf{(b)} The RNN consisted of 100 recurrently connected units which received 32 orientation-tuned inputs and two go-cues that switched from 0 to 1 to initiate the RNN's responses. The target outputs were $\cos$ and $\sin$ of two times the line orientations.}
 \vspace{-10pt}
\label{fig:taskandarchitecture}
\end{figure}

\section{Methods}
\label{sec:methods}

We first describe our standard setup and then the variations.
The training task and network architecture are shown schematically in Figure \ref{fig:taskandarchitecture}. Two arbitrary input orientations ($\theta$ and $\phi$) were presented successively to the RNN with a variable delay (Delay 1) between them. After another variable delay (Delay 2), input go-cues were presented to indicate when to decode the absolute orientations, and in some simulations, their ordinal relationship (see Figure \ref{fig:RNNpsychophysics}), at the output. 
We used 32 orientation-tuned input units whose preferred orientations cover the full 180$^\circ$ range, and modeled their activities after V1 orientation selective cells \citep{Teich2003}. The go-cue inputs were binary 0 and then 1 to initiate the RNN's response. The RNN consisted of 100 fully-connected units. We used the $\cos$ and $\sin$ of $2\theta$ and $2\phi$ as absolute-orientation outputs so that a line oriented at 0$^\circ$ had the same output as a line oriented at 180$^\circ$.
When the ordinal output was included, it was a binary unit for the clockwise/counterclockwise relationship between the two input lines (see Section \ref{sec:results} for more details).

The dynamics $u_i(t)$ of each recurrent unit in the network 
was governed by the standard continuous-time RNN equations:
\begin{align}
	\label{eq:ningpaper_dynamics1}
	\tau \frac{dx_{i}(t)}{dt}   &=  -x_{i}(t) + \sum_{j=1}^{N^{\mathrm{rec}}}W^{\mathrm{rec}}_{ij}u_{j}(t)  + \sum_{k=1}^{N^{\mathrm{in}}}W^{\mathrm{in}}_{ik}I_{k}(t) + b_{i}   \\
	u_{i}(t)  &=  f(x_{i}(t)) + \xi_{i}(t) 
	\label{eq:ningpaper_dynamics2}
\end{align}
for $i=1,\ldots,N^{\mathrm{rec}}$. 
The activity $u_{i}(t)$ of unit $i$ at time $t$ was computed based on $x_{i}(t)$ through a rectified $\tanh$ nonlinearity $f(x) = \max(0,\tanh(x))$.
Each unit received input from other units through recurrent connections with weights determined by the matrix $W^{\mathrm{rec}}$, initialized orthogonally \citep{Saxe2014}.
The units also received input $I(t)$ that entered the RNN through input weights determined by the matrix $W^{\mathrm{in}}$. 
Each unit had two sources of bias: (1) $b_{i}$ which was learned and (2) $\xi_{i}(t)$ which represented noise intrinsic to the network and was taken to be white Gaussian (independently sampled at each time step) with zero mean. 
The network was simulated using the Euler method for $T=500$ timesteps, each of duration $\tau/10$ \citep{ManteSussillo2013}.

To perform the psychophysics task with the RNN, we linearly combined the activity of recurrent units to decode the output $y_{j}(t)$ according to: 
\begin{equation}
	y_{j}(t) = \sum_{i=1}^{N^{\mathrm{rec}}}W^{\mathrm{out}}_{ji}u_{i}(t).
	\label{eq:ningpaper_output}
\end{equation}

We optimized the network parameters $W^{\mathrm{in}}$, $W^{\mathrm{rec}}$, $b$ and $W^{\mathrm{out}}$ to minimize the mean squared error between the target outputs and the network outputs:
\begin{equation}
	E = \frac{1}{MTN^{\mathrm{out}}}\sum_{m,t,j=1}^{M,T,N^{\mathrm{out}}} (y_{j}(t,m) - y^{\mathrm{target}}_{j}(t,m))^{2} 		
	\label{eq:ningpaper_error}
\end{equation}
Parameters were updated with the Hessian-free algorithm~\citep{Martens2011} using mini-batches of size $M=500$ trials. We also varied many aspects of the above standard setup to confirm robustness of our conclusions. Most importantly, we used different output representations, different activation functions, different learning algorithm (Adam, \cite{Kingma2015}), and simultaneous instead of sequential reporting for the two orientations.
The results reported below remained consistent in the above variants.

\section{Results}
\label{sec:results}

\begin{figure}[!h]
\vspace{-.05in}
\begin{subfigure}{0em}\phantomsubcaption{}\label{fig:toriA}\end{subfigure}
\begin{subfigure}{0em}\phantomsubcaption{}\label{fig:toriB}\end{subfigure}
\begin{subfigure}{0em}\phantomsubcaption{}\label{fig:toriC}\end{subfigure}
\begin{subfigure}{0em}\phantomsubcaption{}\label{fig:toriD}\end{subfigure}
\begin{subfigure}{0em}\phantomsubcaption{}\label{fig:toriE}\end{subfigure}
\begin{subfigure}{0em}\phantomsubcaption{}\label{fig:toriF}\end{subfigure}
\begin{subfigure}{0em}\phantomsubcaption{}\label{fig:toriG}\end{subfigure}
\begin{subfigure}{0em}\phantomsubcaption{}\label{fig:toriH}\end{subfigure}
\centering
\includegraphics[trim={0cm 11cm 2cm 0cm},clip,keepaspectratio,width=1.0\linewidth]{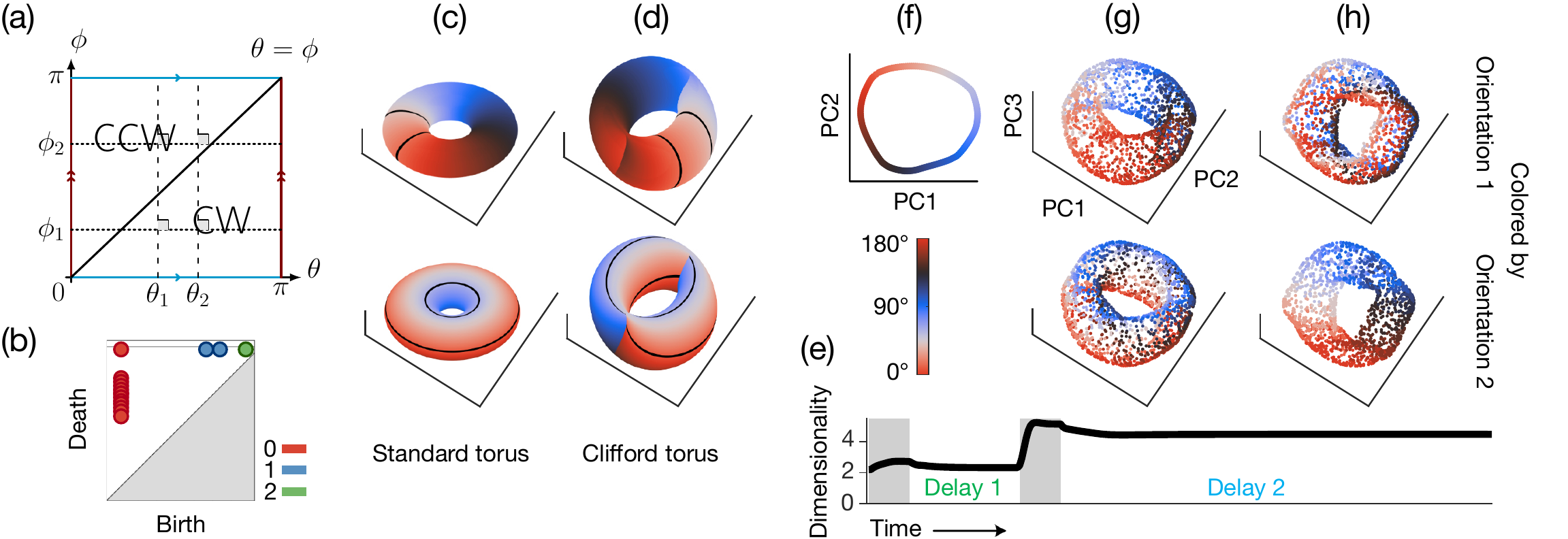}
\caption[Standard and Clifford tori]{Standard vs. Clifford tori and memory geometry in the RNN. \textbf{(a)} Fundamental polygon of both types of tori. Identifying the blue edges together and the red edges together results in a standard torus if done in 3D ambient space, while it results in a Clifford torus if done in 4D ambient space. \textbf{(b)} Persistence diagram for both types of tori is the same. \textbf{(c)} Example of a standard torus colored by the two angular parameters. The black curves show rings obtained by fixing one of the two angular parameters, e.g. rings corresponding to the dashed vertical line at $\theta_1$ or dotted horizontal line at $\phi_2$ in panel (a). \textbf{(d)} Example 3D projections of a Clifford torus colored by the two angular parameters. \textbf{(e)} Dimensionality of the RNN activity over time and across all trials. Shaded regions denote the intervals when orientation 1 and orientation 2 were presented. The dimensionality, quantified by the participation ratio \citep{Rajan2010,Abbott2011}, is near 2 after the presentation of the first orientation and near 4 after both orientations have been presented. \textbf{(f)} The RNN activity at the end of the first delay period is shown projected onto the first two principal components (2 PCs explained 92\% of variance). The first angular variable is stored around a ring.  \textbf{(g-h)} The RNN activity is shown at two different timepoints after the presentation of the second orientation, after projecting onto the first three principal components (3/4 PCs explained 74\%/93\% of variance at the end of Delay 2). ``Switching'' of the subspaces encoding orientation 1 and orientation 2 between (g) and (h) is consistent only with the Clifford torus geometry. } 
 \vspace{-10pt}
\label{fig:tori}
\end{figure}

\paragraph{Low-dimensional manifolds of recurrent-unit activities}  To understand how neural circuits may represent multiple items of continuous values, we trained RNNs to remember two successively presented input orientations and decoded them later (Figure \ref{fig:taskandarchitecture}). After training we tested the RNNs using various combinations of the two input orientations and recorded the activities of the recurrent units over time. We then applied PCA to the activities to examine their low-dimensional structure. When interrogating the RNN with various line 1 stimuli, the activity manifold settles into a ring attractor during Delay 1 as expected (Figure \ref{fig:toriF}). 
We next investigated the activity manifold after the presentation of both lines. 

To see what results may be expected, consider the square region of Figure \ref{fig:toriA} which represents all possible combinations of the two input orientations in a trial, with each axis covering the full 180$^\circ$ range of each orientation. Since each orientation is periodic, each pair of opposite edges of the square should be identified. If we do the two identifications in $\mathbb{R}^3$,
we need to distort the square to form a standard torus (the surface of a donut) shown in Figure \ref{fig:toriC}. 
This distortion, however, has undesirable consequences for the task as the two orientations are treated very differently. The first orientation could be stored on the toroidal rings (as shown in the top panel of Figure \ref{fig:toriC}) and the second orientation could be stored on the poloidal rings (Figure \ref{fig:toriC} bottom panel), or vice versa. Since toroidal rings have different circumferences, they represent the same orientation differently when the other orientation changes. In contrast, the task requires equally accurate recall of both orientations. When we consider ordinal output later, we will show yet another undesirable consequence of the distortion.

\begin{figure}[!h]
    \vspace{-.05in}
    \begin{subfigure}{0em} \phantomsubcaption{} \label{fig:testsA} \end{subfigure}
    \begin{subfigure}{0em} \phantomsubcaption{} \label{fig:testsB} \end{subfigure} 
    \begin{subfigure}{0em} \phantomsubcaption{} \label{fig:testsC} \end{subfigure} 
    \begin{subfigure}{0em} \phantomsubcaption{} \label{fig:testsD} \end{subfigure} 
    \begin{subfigure}{0em} \phantomsubcaption{} \label{fig:testsE} \end{subfigure} 
    \begin{subfigure}{0em} \phantomsubcaption{} \label{fig:testsF} \end{subfigure} 
    \begin{subfigure}{0em} \phantomsubcaption{} \label{fig:testsG} \end{subfigure} 
    \begin{subfigure}{0em} \phantomsubcaption{} \label{fig:testsH} \end{subfigure} 
    \begin{subfigure}{0em} \phantomsubcaption{} \label{fig:testsI} \end{subfigure} 
    \begin{subfigure}{0em} \phantomsubcaption{} \label{fig:testsJ} \end{subfigure} 
    \begin{subfigure}{0em} \phantomsubcaption{} \label{fig:testsK} \end{subfigure} 
    \begin{subfigure}{0em} \phantomsubcaption{} \label{fig:testsL} \end{subfigure} 
    \begin{subfigure}{0em} \phantomsubcaption{} \label{fig:testsM} \end{subfigure} 
    \begin{subfigure}{0em} \phantomsubcaption{} \label{fig:testsN} \end{subfigure} 
    \begin{subfigure}{0em} \phantomsubcaption{} \label{fig:testsO} \end{subfigure} 
    \begin{subfigure}{0em} \phantomsubcaption{} \label{fig:testsP} \end{subfigure} 
    \begin{subfigure}{0em} \phantomsubcaption{} \label{fig:testsQ} \end{subfigure} 
    \begin{subfigure}{0em} \phantomsubcaption{} \label{fig:testsR} \end{subfigure} 
    \centering
    \includegraphics[trim={0cm 7.3cm 0.5cm 0cm},clip,keepaspectratio,width=1\linewidth]{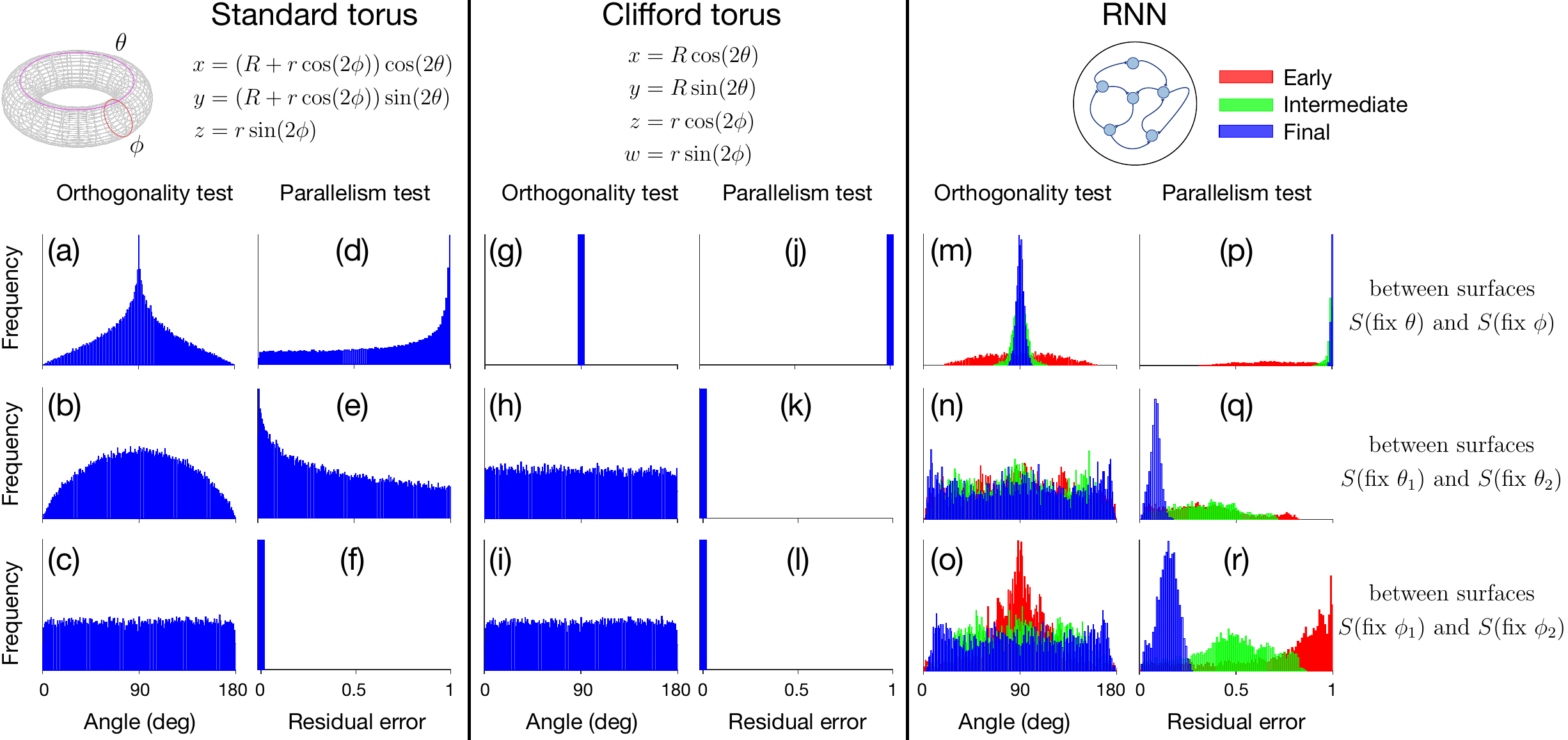}
	\caption{Orthogonality and parallelism tests for \textbf{(a-f)} a standard torus, \textbf{(g-l)} Clifford torus, and \textbf{(m-r)} trained RNN. The scales of the y-axes are chosen to make the shapes of the distributions easier to see and are not all the same. The Clifford torus and RNN store the two orientations in orthogonal subspaces whereas the standard torus does not (see first row). See text for explanations of the orthogonality and parallelism tests. Comparing the second and third rows, we see the standard torus treats each angle asymmetrically, e.g. the orthogonality tests for the standard torus are not the same for both angles. Similarly, the parallelism tests are also not the same for both angles. In contrast, the Clifford torus and RNN store the two angles in a similar manner.}
	 \vspace{-10pt}
\label{fig:tests}
\end{figure}

Alternatively, we can identify the two sets of opposite edges of the square in $\mathbb{R}^4$ without distorting the square at all, to produce what is known as the Clifford torus. It is a two-dimensional manifold embedded in $\mathbb{R}^4$ defined as the Cartesian product of two circles each embedded in $\mathbb{R}^2$ \citep{cliffortoruswiki}. Although topologically equivalent to a standard torus (Figure \ref{fig:toriA}) a Clifford torus is intrinsically flat and treats the two orientations, and different values of the same orientation, equally, as demanded by the task \citep{Weeks2019}. Intuitively, one cannot fit the product of two circles in $\mathbb{R}^3$ without distortion (because each 1D circle requires a 2D space to be embedded in) but can do so in $\mathbb{R}^4$.

We now show our analyses for differentiating these two geometric objects.
When we first plotted the activity manifold in the space of the first 3 PCs, it looked like a standard torus with the two orientations represented along the toroidal and poloidal rings, respectively (Figure \ref{fig:toriG}). However, when we followed the manifold over time, we noticed that the manifold appeared to run though itself such that at a later time, the two orientations represented by the toroidal and poloidal rings swapped (Figure \ref{fig:toriH}). Since the set of differential equations governing the system is autonomous when these activities are recorded (no time dependent inputs), the solution should be unique, suggesting that the actual manifold is embedded in a higher-dimensional space without running through itself. We confirmed that the embedding space was indeed roughly four dimensional \citep{Rajan2010,Abbott2011,Mazzucato2016,Gao2017}: the top 3 and 4 PCs explained 74\% and 93\% of the variance at the end of the second delay period (Figure \ref{fig:toriE}).

Next, standard and Clifford tori make different predictions about the geometry (orthogonality and parallelism) of the subspaces formed by fixing one orientation while varying the other. We tested these predictions.
For convenience, we denote the subspace $\mathbf{S}_{\theta}$ when we fixed the first orientation at $\theta$ and varied the second orientation, and $\mathbf{S}_{\phi}$ when we fixed the second orientation at $\phi$ and varied the first orientation. For example, the two black curves on the torus in Figure \ref{fig:toriD} (e.g., top panel) could correspond to $\mathbf{S}_{\theta=100^{\circ}}$ and $\mathbf{S}_{\theta=125^{\circ}}$. A Clifford torus predicts that different $\mathbf{S}_{\theta}$s (or different $\mathbf{S}_{\phi}$s) should be parallel to each other (Figure \ref{fig:testsH}-\ref{fig:testsI}) whereas $\mathbf{S}_{\theta}$s and $\mathbf{S}_{\phi}$s should be orthogonal to each other\footnote{Two (vector) subspaces $A$ and $B$ of an inner product space $V$ are called orthogonal subspaces if each vector in $A$ is orthogonal to each vector in $B$.} (Figure \ref{fig:testsG}). These predictions clearly do not hold for a regular torus (Figures \ref{fig:testsA}-\ref{fig:testsC}).

To test the orthogonality prediction, we calculated the angles between the first two PCs of $\mathbf{S}_{\theta}$'s and the first two PCs of $\mathbf{S}_{\phi}$'s. These angles distributed around 90 degrees (Figure \ref{fig:testsM}) for the RNN, supporting the Clifford torus hypothesis. As a control, we also calculated the angles between the PCs spanning different $\mathbf{S}_{\theta}$s (or similarly, different $\mathbf{S}_{\phi}$s); they distributed near uniformly in the interval $[0, 180)$ degrees (Figure \ref{fig:testsN}-\ref{fig:testsO}).

To test the parallelism prediction, we calculated the residuals of reconstructing one $\mathbf{S}_{\theta}$'s PCs using another $\mathbf{S}_{\theta}$'s PCs, or one $\mathbf{S}_{\phi}$'s PCs using another $\mathbf{S}_{\phi}$'s PCs.
For near-parallel subspaces, we expect these residuals to be close to 0 whereas for orthogonal subspaces they should be close to 1 (for unit-length PCs). The results in Figures \ref{fig:testsQ}-\ref{fig:testsR} show that the residuals distributed near 0, again supporting the Clifford torus hypothesis. As a comparison, we also calculated the residuals of reconstructing an $\mathbf{S}_{\theta}$'s PCs using an $\mathbf{S}_{\phi}$'s PCs and vice versa. As expected, the residuals distributed near 1 for the Clifford torus (Figure \ref{fig:testsJ}) and RNN (Figure \ref{fig:testsP}).

The above analyses indicate that the low-dimensional manifold of the recurrent unit activities, after the presentation of both orientations, resembled a Clifford torus more than a standard torus.
We then examined how the Clifford-torus-like structure emerged during the training process. The colored panels in Figure \ref{fig:tests} show the orthogonality and parallelism results for the RNN at three stages during training. Early in training, the results did not look like those of a Clifford torus (red) but they evolved with further training (green) until the final geometry of the memory representation resembled a Clifford torus (blue).
We posit that storing the two orientations in nearly orthogonal subspaces makes the memories more robust to noise.

To check whether formation of the Clifford tori is specific to our choice of unit activation function, noise level and learning algorithm, we experimented with a different output representation for orientations (same as the input), other activation functions ($\tanh$ and ReLU), various noise levels, and a different learning algorithm (Adam, \cite{Kingma2015}). In all of the cases, we obtained similar results which suggests the storage of the two orientations on near-orthogonal subspaces was a rather general strategy for solving the task.

Topological data analysis (TDA) has been applied to study activity manifolds in both real and artificial neural networks. As we noted previously, Clifford and standard tori are topologically equivalent; consequently, TDA cannot tell them apart. In contrast, the orthogonality and parallelism tests we developed can. Although TDA is a powerful and useful tool, our work demonstrates the importance of going beyond TDA, allowing us to answer new questions about the geometry of working memory.


\paragraph{Analysis of Network Connectivity.}

\begin{figure}[!ht]
\vspace{-.05in}
\begin{subfigure}{0em} \phantomsubcaption{} \label{fig:tuningandweights_noordinaloutputA} \end{subfigure}
\begin{subfigure}{0em} \phantomsubcaption{} \label{fig:tuningandweights_noordinaloutputB} \end{subfigure} 
\begin{subfigure}{0em} \phantomsubcaption{} \label{fig:tuningandweights_noordinaloutputC} \end{subfigure} 
\begin{subfigure}{0em} \phantomsubcaption{} \label{fig:tuningandweights_noordinaloutputD} \end{subfigure} 
\begin{subfigure}{0em} \phantomsubcaption{} \label{fig:tuningandweights_noordinaloutputE} \end{subfigure} 
\begin{subfigure}{0em} \phantomsubcaption{} \label{fig:tuningandweights_noordinaloutputF} \end{subfigure} 
\centering
\includegraphics[trim={0cm 4.7cm 2.4cm 0cm},clip,keepaspectratio,width=1\linewidth]{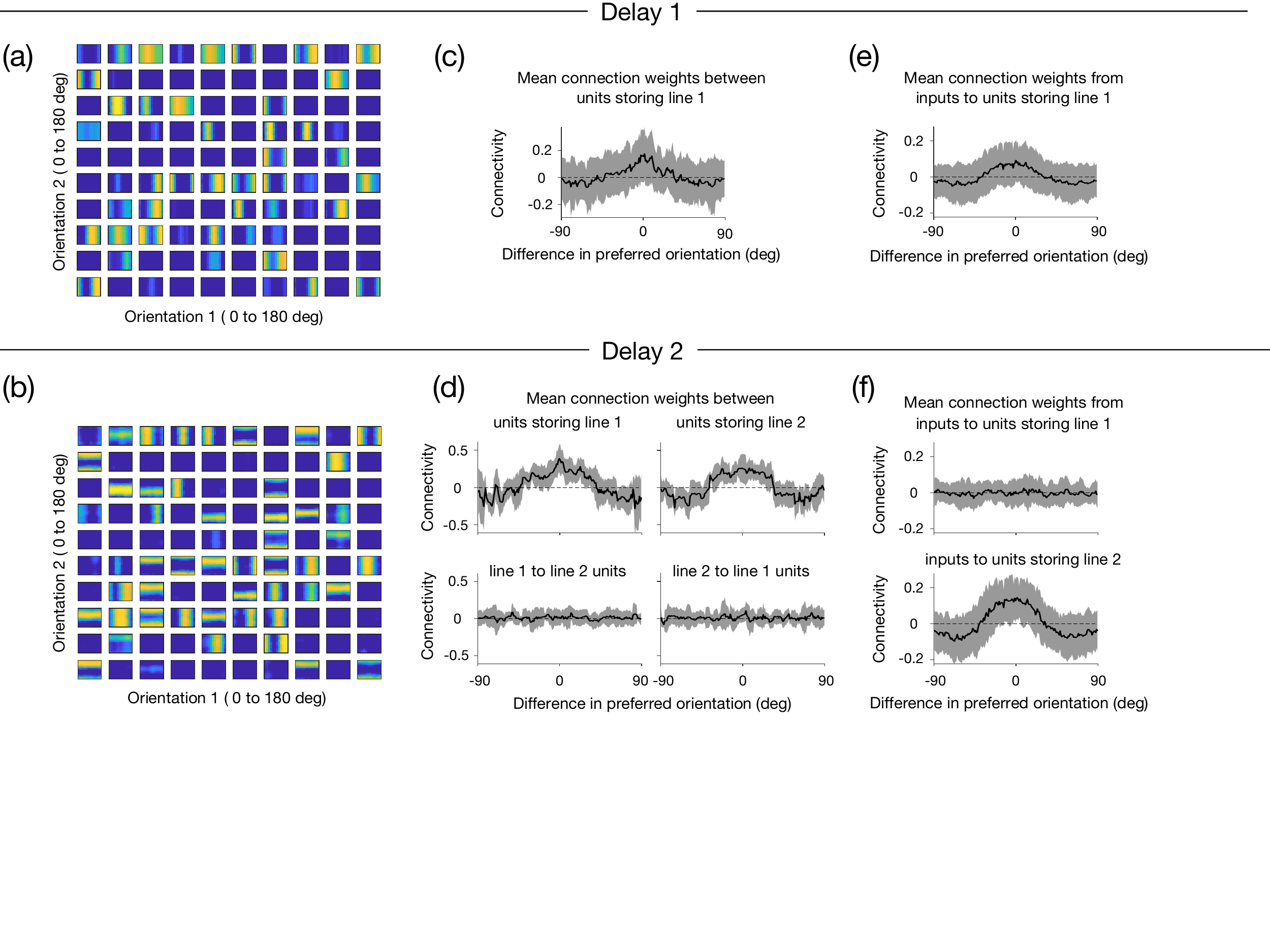}
 \caption{
 Tuning (leftmost column) and connectivity (other columns) in a trained RNN during the first delay period after the presentation of the first line (top row, delay 1) and during the second delay period after the presentation of the second line (bottom row, delay 2). \textbf{(a-b)} Joint tuning of all 100 recurrent units to the two orientations are shown at their steady state values, for the two delay periods. The activity of each unit is shown as a function of line 1 orientation (x-axis) and line 2 orientation (y-axis), with yellow indicating high activity and blue indicating no activity. \textbf{(c-d)} Mean connection weights between the recurrent units as a function of their preferred-orientation difference during the two delay periods. The classic local-excitation/global-inhibition connectivity motif developed between units tuned to the same line (panel c and the top row of panel d), but not between units tuned to different lines (bottom row of panel d).  \textbf{(e-f)} Mean connection weights from the inputs to the recurrent units as a function of their preferred-orientation difference during the two delay periods. The input connections to the recurrent units tuned to line 1 in the first delay period had a local-excitation/global-inhibition pattern (panel e). This pattern disappeared during the second delay period (top row of panel f) because recurrent units changed their tuning to line 1. However, the same connectivity pattern emerged from the inputs to the recurrent units tuned to line 2 (bottom row of panel f).  In panels c-f, error bars show one standard deviation.}
 \vspace{-10pt}
\label{fig:tuningandweights_noordinaloutput}
\end{figure}

We next looked at the tuning properties and connectivity motifs in the trained network to understand how they could support the activity structure described above. The neural activities of all 100 units in the RNN are shown in Figures \ref{fig:tuningandweights_noordinaloutputA} and \ref{fig:tuningandweights_noordinaloutputB}. Each small heatmap represents one recurrent unit's activity as a function of the first orientation (x-axis) and the second orientation (y-axis), with yellow indicating high activity and blue indicating no activity. For each unit, the tuning is shown as orientation 1 and orientation 2 vary between 0 and 180 degrees.

The joint tuning is shown at the steady state after line 1 presentation (Delay 1, Figure \ref{fig:tuningandweights_noordinaloutputA}) and after line 2 presentation (Delay 2, Figure \ref{fig:tuningandweights_noordinaloutputB}). We found that at a given time, most units were tuned to either the first or second orientation (vertical or horizontal stripes, respectively), consistent with the near-orthogonal subspaces for the two orientations.

We wondered what connectivity patterns among the units could support the attractors we found above. We first divided the units into two classes according to which line they were tuned to, and determined each unit's preferred orientation that yields maximum activity. Then for each class of units, we plotted the mean connection strength between the units as a function of the difference in their preferred orientations, and found a highly structured connectivity pattern of local excitation and global inhibition as shown in Figures \ref{fig:tuningandweights_noordinaloutputC} and \ref{fig:tuningandweights_noordinaloutputD}. 
Specifically, within a class, units that had similar preferred orientations were connected through positive weights, and units with more different preferred orientations were connected through negative weights. This local-excitation/global-inhibition connectivity pattern is well known to generate ring attractors for representing periodic variables such as orientation and heading direction \citep{Somers1995,Zhang1996,Teich2003,Cueva2019}. As a control, we verified that this connectivity did not exist between the two classes of units (Figure \ref{fig:tuningandweights_noordinaloutputD} bottom row), which may support the near orthogonal subspaces for the two orientations.

\paragraph{Dynamic coding.}

We noticed that the tuning of some units in the RNN changed over time (Figures \ref{fig:tuningandweights_noordinaloutputA} and \ref{fig:tuningandweights_noordinaloutputB}). For example, a unit that stored information about line 1 during the first delay period did not always continue to store information about line 1 during the second delay period. Do these changes in tuning confer some computational benefit? In particular, both input orientations entered the RNN through the exact same set of fixed connection weights, namely $W^{\mathrm{in}}$ in Equation (\ref{eq:ningpaper_dynamics1}); could the tuning change help prevent line 2 orientation from overwriting the memory of line 1 orientation?    

To help answer this question, we plotted the mean connection weights from the 32 inputs to the 100 recurrent units as a function of the difference between their preferred orientations, in Figures \ref{fig:tuningandweights_noordinaloutputE} and \ref{fig:tuningandweights_noordinaloutputF}. We found that during the first delay period, there is a local-excitation/global-inhibition connectivity pattern from the inputs to the recurrent units tuned to line 1. However, during the second delay period, this pattern disappeared as the recurrent units changed their tuning \citep{Libby2021}. Meanwhile, a similar local-excitation/global-inhibition connectivity pattern emerged from the inputs to the RNN units tuned to line 2. Thus, the tuning change appeared to both ``move'' the working memory of line 1 orientation and ``hide'' it from the inputs, while ``making room'' for the network to encode new sensory information of line 2.


\paragraph{Reproducing Psychophysics Results of Ding et al.} 

With the above understanding of how an RNN stores two orientations, we now present our RNN implementation of Ding et al.'s \citep{Ding2017} decoding scheme that explains their psychophysical data. In their experiment, when subjects rotated markers to report two remembered, absolute orientations, they also implicitly indicated the ordinal relationship between the orientations. The key findings were that the two reported absolute orientations in a trial were correlated and that the second line repelled the first line (backward aftereffect) as much as the first line repelled the second line (forward repulsion). To explain the results, Ding et al.\ used the ordinal relationship as a Bayesian prior to constrain the decoding of the absolute orientations. They argued that such a high-to-low-level decoding scheme is advantageous when the working memory is noisy. 
We thus hypothesized that if human behavior in this task relies on making higher-level, more categorical judgments and these judgements are prioritized due to noisy working memory, then RNNs trained to make categorical judgements in the presence of noise should more accurately capture these behavioral patterns. To test this, we trained different RNNs to solve this task and then compared the behavior of humans and RNNs. Specifically, we trained RNNs either with or without an ordinal output as an additional optimization goal, and with or without injecting noise into the RNN, particularly during Delay 2 when both orientations are stored.  

\begin{figure}[!h]
\vspace{-.05in}
\begin{subfigure}{0em} \phantomsubcaption{} \label{fig:RNNpsychophysicsA} \end{subfigure}
\begin{subfigure}{0em} \phantomsubcaption{} \label{fig:RNNpsychophysicsB} \end{subfigure} 
\begin{subfigure}{0em} \phantomsubcaption{} \label{fig:RNNpsychophysicsC} \end{subfigure} 
\begin{subfigure}{0em} \phantomsubcaption{} \label{fig:RNNpsychophysicsD} \end{subfigure}
\begin{subfigure}{0em} \phantomsubcaption{} \label{fig:RNNpsychophysicsE} \end{subfigure} 
 \centering
  \includegraphics[trim={0cm 11cm 3.4cm 0cm},clip,keepaspectratio,width=1\linewidth]{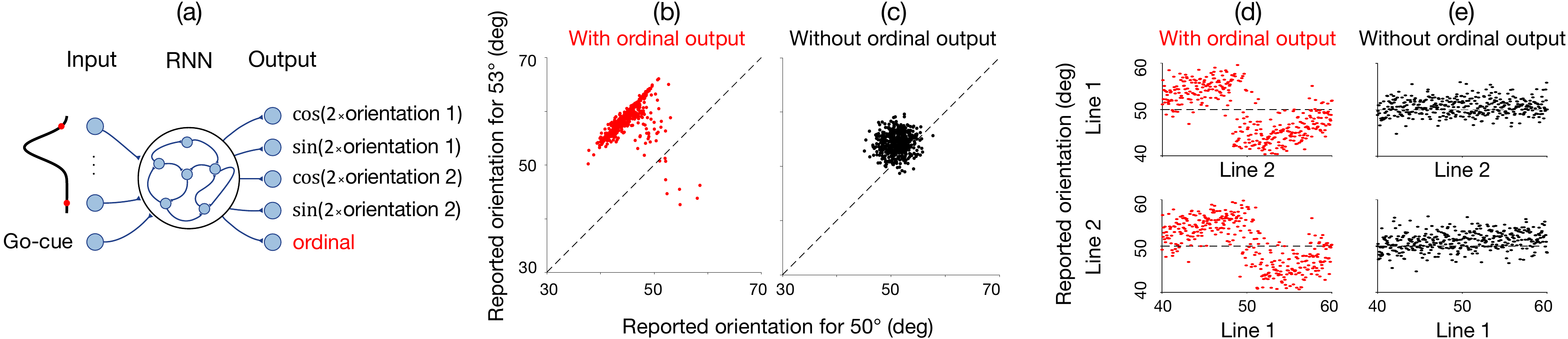}
 \caption[RNN orientation outputs are correlated]{Human behavior was not captured by all RNNs. \textbf{(a)} RNN optimized in the presence of memory noise to report both absolute orientations and their ordinal relationship, i.e. whether the second orientation is a clockwise or counterclockwise rotation from the first. The RNN consisted of 100 recurrently connected units which received 32 orientation-tuned inputs and a go-cue that switched from 0 to 1 to initiate the RNN's responses. The target outputs were $\cos$ and $\sin$ of two times the line orientations, and a binary indicator for their ordinal relationship. To remove potential biases due to an asymmetric motor response, and probe the sufficiency of memory noise plus the ordinal constraint to explain the behavioral results, all outputs were reported simultaneously. However, our results held in both sequential and simultaneous settings.   \textbf{(b)}  After training, the network in (a) was probed with line orientations of $50^{\circ}$ and $53^{\circ}$ as in \cite{Ding2017}. The RNN's outputs of the two lines' orientations were correlated and repelled from the diagonal line ($\theta=\phi$) as observed in the psychophysics results of Ding et al. \textbf{(c)} 
 In contrast, RNNs trained without this inductive bias (i.e. without the ordinal output) did not recapitulate human behavior. These  results  are  consistent  with  the  hypothesis  that  humans  use  higher-level strategies to solve this task. 
  \textbf{(d-e)} Forward and backward aftereffect. Each dot indicates the result from a separate trial where the orientations of line 1 and line 2 were varied. In this example the target output for the RNN is always $50^{\circ}$ (dotted black line). For the RNN with the ordinal output (d), its reported orientations were biased by the orientation of the other line in way that is consistent with the ordinal memory, and with the human psychophysics of Ding et al. Notice the RNN's report for the orientation of line 1 was affected by line 2 which appeared \emph{afterwards}, a result not predicted by a standard feedforward network (top panel of d). In contrast, for the RNN without the ordinal output the reported orientations were not affected by the orientation of the other line (e).}
 \vspace{-10pt}
 \label{fig:RNNpsychophysics}
 \end{figure}

To test the hypotheses, we trained three main versions of the RNN. To remove the confound of an asymmetric motor response, and probe the sufficiency of memory noise plus the ordinal constraint to explain the behavioral results, we trained RNNs to report all orientations simultaneously. However, our findings hold for RNNs with both sequential and simultaneous reports.
Version 1 was the same as the RNN above which was only trained to output the two absolute orientations, but not their ordinal relationship.
We probed the trained network with line 1 and line 2 oriented at $50^{\circ}$ and $53^{\circ}$, as in Ding et al.'s experiment, with noise added to the firing rates to create trial-to-trial variability. The network generated a distribution of predictions, spherically centered around the true values as shown in Figure \ref{fig:RNNpsychophysicsC}, quite different from the actual data (see panel A of Figure 3 in \cite{Ding2017}). A spherical distribution was obtained for all RNNs with no ordinal output, regardless of whether they were trained with noise or without.  

To examine the effect of the ordinal relationship between the lines, we incorporated an additional ordinal output in version 2 as shown in Figure \ref{fig:RNNpsychophysicsA}. This output was +1 if the orientation of the second line was clockwise from the first line, and -1 if counterclockwise.  For this version, we did not add any firing rate noise $\xi_{i}(t)$ during training.  
When it was tested on inputs of $50^{\circ}$ and $53^{\circ}$ (again, with a small amount of noise added to the firing rates to create trial-to-trial variability) the output distribution was, once more, spherical and centered around the true values as in Figure \ref{fig:RNNpsychophysicsC}. Even though the ordinal relationship was stored by the RNN, it was not used to constrain the decoding of the two absolute orientations.

To reproduce the psychophysics results, we needed to force the RNN to use the ordinal memory by introducing noise into the network during training, making the continuous orientation memories less stable than the binary ordinal memory. Version 3 of the network included the ordinal output as in version 2, but additionally, noise was injected into the firing rates of the network during training via $\xi_{i}(t)$ in Equation (\ref{eq:ningpaper_dynamics2}). 
As shown in Figure \ref{fig:RNNpsychophysicsB}, in this case the RNN outputs of the two absolute orientations are correlated with, and repelled from, each other, and display similar forward and backward aftereffects (Figure \ref{fig:RNNpsychophysicsD}-\ref{fig:RNNpsychophysicsE}) in agreement with the psychophysics results of \cite{Ding2017}. Intuitively, the noise injection made it difficult to get the ordinal output correct when the two absolute orientations were nearly identical (i.e., near the diagonal line in the joint space of Figure \ref{fig:toriA}). The ordinal training, then, must force a coordinated shift of the absolute orientations away from the diagonal line to produce the correlation and repulsion.

Version 3 is also consistent with the original interpretation of \cite{Ding2017} that the retrospective high-to-low-level decoding is advantageous when the decoding occurs in noisy working memory. Interestingly, we did not have to train the network on the higher-level, ordinal relationship before the lower-level, absolute orientations. Instead, training all these outputs together produced the result that the ordinal relationship constrained the absolute orientations, but not vice versa. This is likely because the ordinal relationship is categorical and thus easier to learn and maintain in the noisy recurrent units, compared with the continuous, absolute orientations. 

We also analyzed the memory geometry of the RNN that explains the psychophysics data (i.e., the RNN with the ordinal output and noise added). We found that the activity manifold after the presentation of the second orientation appeared to be a distorted Clifford torus: the subspaces for the different orientations were still nearly orthogonal but not as close to orthogonal as in Figure \ref{fig:tests}, and the subspaces for two values of the same orientation were no longer parallel. 
This distortion must be for accommodating the correlation and repulsion between the two lines. Interestingly, the manifold still resembled a Clifford torus more than a standard torus. The reason is likely that the metric structure of the standard torus is also undesirable for the ordinal output. Consider the diagonal line in Figure \ref{fig:toriA}, which is the decision boundary for the ordinal output. When the opposite sides of the square are identified in 3D to form the standard torus, the decision boundary becomes a curve not contained in a plane (this can be shown by demonstrating that the torsion of the curve is not 0 everywhere). Consequently, the two opposite ordinal outcomes are not linearly separable. In contrast, when the opposite sides of the square are identified in 4D to form a Clifford torus, the decision boundary does not change and the ordinal outcomes remain linearly separable.  

To summarize, there are two requirements during training for reproducing Ding et al's psychophysical data: memory noise and ordinal output. We believe this is an interesting finding because previous literature may have created the (improper) expectation that training neural networks to perform a task well automatically produces brain-like computation \citep{Yamins2014, Cadieu2014, Yamins2016}. In contrast, we found that simply training a network to accurately report absolute orientations does not generate a brain-like system. We also have to consider factors that constrain neural computation \citep{Sussillo2015}, such as memory noise and ordinal output.

In addition, our network model allowed us to explicitly probe the hypothesis that memory noise coupled with the ordinal constraint was sufficient to reproduce the psychophysics behavior. We trained models where noise was only injected during the second delay period when both memories must be stored, demonstrating that motor and sensory noise were not necessary to explain human behavior on this task.

\section{Discussion and Conclusions}

We investigated how neural circuits might store multiple items of continuous values in working memory by training RNNs to report orientations of two previously presented lines. We analyzed the dimensionality of the embedding space, and the relationships between subspaces of the recurrent-unit activities, and found that after the presentation of the second line, the low-dimensional activity manifold resembled a Clifford torus more than a standard torus. 
In order to disambiguate these two memory geometries we could not rely on tools from topological data analysis (Figure \ref{fig:toriB}), as the Clifford and standard tori are topologically equivalent. Therefore we introduced the orthogonality and parallelism tests to disambiguate these two possibilities and reveal the geometry of working memory. We argued that the Clifford torus better matches the task demands because it treats the two line orientations equally, and stores them in orthogonal subspaces to avoid interference. We then examined the connections among the units, and found that for units tuned to the same line (first or second), those preferring similar orientations excited each other and those preferring different orientations inhibited each other. There were thus two sets of locally-excitatory/globally-inhibitory connectivity patterns, one for each line. Such connectivity patterns did not exist between units tuned to the different lines. Therefore, the overall connectivity supported the Clifford-torus-like activity manifold that stored the two line orientations in nearly orthogonal subspaces. 

We further found that the recurrent units changed their orientation preferences over time. By analyzing the connectivity patterns from the input units to the recurrent units at different times, we provided evidence that this dynamic
code appeared to safeguard the memory of the first orientation by ``moving'' it to a different subspace while the network encoded the second orientation, resolving the conflict between preserving the memory of old information and encoding new sensory input. 
This ``dynamic memory'' mechanism must be generally required when a working memory system receives a sequence of inputs through the same sensory channel.

Finally, we tested the hypothesis that visual perception in the task of \cite{Ding2017} is actually high-to-low-level decoding in noisy working memory, with higher-level features constraining the decoding of lower-level features.
Specifically, we simulated Ding et al.'s task of reporting the absolute orientations of, and the ordinal relationship between, two successively flashed lines. We demonstrated that noise injected into the recurrent units naturally led to the higher-level ordinal relationship constraining the decoding of the lower-level absolute orientations to reproduce the key psychophysical findings of \cite{Ding2017}.
Noise was necessary; without it the RNN did not use the higher-level ordinal memory to constrain the remembered values of the two orientations.

Although our learning tasks were very simple, these tasks and their many variants have been widely used in visual psychophysics literature \citep{Jazayeri2007, Stevenson2009, Zhang2009, Brady2011, Ma2014, Bae2015, GreenSwets1966}. Our study thus provides a thorough understanding of a neural network implementation for a major class of psychophysical paradigms, from the behavioral level to the activity and connectivity levels. Additionally, although high-to-low-level decoding is a property of many auto-encoders, standard feedforward networks do not model the effect of noisy working memory on decoding hierarchy, which was necessary for reproducing human behavior in our RNN model. Our work may thus inspire future efforts on understanding visual decoding hierarchy from a machine learning perspective.

\acks{
This work was supported by NSF grant 1754211, NSF NeuroNex Award DBI-1707398, and the Gatsby Charitable Foundation. 
Simulations were partially performed using computing facilities on {CloudLab} \citep{cloudlab2019}.}

\bibliography{bibliography.bib}

\end{document}